\documentclass[useAMS,usenatbib]{mn2e} 
\usepackage{aas_macros}
\usepackage{graphics}
\usepackage[pdftex]{graphicx}
\usepackage{epstopdf}
\usepackage{epsfig}  
\usepackage{natbib} 
\usepackage{float}
\usepackage{amsmath}
\usepackage{times}
\usepackage[varg]{txfonts}
\usepackage{verbatim} 
\usepackage{multirow,bigdelim} 

\newcommand{\bl}{\boldsymbol}
\newcommand{\tm}{\textrm}

\newcommand{\hMpc}{{\ifmmode{h^{-1}{\rm Mpc}}\else{$h^{-1}$Mpc}\fi}}
\newcommand{\hkpc}{{\ifmmode{h^{-1}{\rm kpc}}\else{$h^{-1}$kpc}\fi}}
\newcommand{\hMsun}{{\ifmmode{h^{-1}{\rm {M_{\odot}}}}\else{$h^{-1}{\rm{M_{\odot}}}$}\fi}}

\def\lesssim{\mathrel{\hbox{\rlap{\hbox{\lower4pt\hbox{$\sim$}}}\hbox{$<$}}}}
\def\gtrsim{\mathrel{\hbox{\rlap{\hbox{\lower4pt\hbox{$\sim$}}}\hbox{$>$}}}}

\title[Constrained simulations from peculiar velocities]
      {Reconstructing cosmological initial conditions from galaxy peculiar velocities. III. Constrained simulations}
\author[Doumler et al.] 
{Timur Doumler$^{1,2}$, Stefan Gottl\"ober$^{2}$, Yehuda Hoffman$^{3}$, and H\'el\`ene Courtois$^{1}$ \\
  $^1$Universit\'e Lyon 1, CNRS/IN2P3, Institut de Physique Nucl\'eaire, 69622 Villeurbanne, Lyon, France\\
  $^2$Leibniz-Institut f\"ur Astrophysik Potsdam, An der Sternwarte 16, 14482 Potsdam, Germany\\
  $^3$Racah Institute of Physics, Hebrew University, Jerusalem 91904, Israel\\
  }

\begin{document}

\date{}

\pagerange{\pageref{firstpage}--\pageref{lastpage}} \pubyear{2012}

\maketitle

\label{firstpage}


\begin{abstract}

In previous works we proposed the Reverse Zeldovich Approximation (RZA) method, which can be used to estimate the cosmological initial conditions underlying the galaxy distribution in the Local Universe using peculiar velocity data. In this paper, we apply the technique to run constrained cosmological simulations from the RZA-reconstructed initial conditions, designed to reproduce the large-scale structure of the Local Universe. We test the method with mock peculiar velocity catalogues extracted from a reference simulation. We first reconstruct the initial conditions of this reference simulation using the mock data, and then run the reconstructed initial conditions forward in time until $z=0$. 
We compare the resulting constrained simulations with the original simulation at $z=0$ to test the accuracy of this method. We also compare them with constrained simulations run from the mock data without the addition of RZA, i.e.\ using only the currently established constrained realizations (CR) method. Our re-simulations are able to correctly recover the evolution of the large-scale structure underlying the data. The results show that the addition of RZA to the CR method significantly improves both the reconstruction of the initial conditions and the accuracy of the obtained constrained resimulations. Haloes from the original simulation are recovered in the re-simulations with an average accuracy of $\approx 2$ Mpc/h on their position and a factor of 2 in mass, down to haloes with a mass of $\approx 10^{14} M_\odot/h$. In comparison, without RZA the re-simulations recover only the most massive haloes with masses of $\approx 5\cdot 10^{14} M_\odot/h$ and higher, and with a systematic shift on their position of about $\approx 10\; \textrm{Mpc}/h$ due to the cosmic displacement field. We show that with the additional Lagrangian reconstruction step introduced by the RZA, this shift can be removed.

\end{abstract}



\section{Introduction}
\label{sec:introduction}

While numerical simulations have developed into a cornerstone of studying the large-scale structure (LSS) of the Universe, there is still a long way to go towards reconciling the predictions drawn from such cosmological simulations with observational data. On the one hand, the general properties of large-scale matter clustering and interacting are now very well understood. This process is typically simulated by generating a random realization of the primordial density fluctuations of the  Universe, and then integrating it forward in time using $N$-body techniques and the physics defined by the concordance $\Lambda$CDM model. These simulations produce model universes which are statistically in good agreement with the observed LSS. On the other hand, the region best studied observationally -- the Local Universe -- shows many properties that cannot be directly modelled with such random realizations. In order to explain the properties and dynamics of the Local Group, i.e.\ the Milky Way and Andromeda galaxies and their satellites, one has to study its formation history, which seems to be tightly connected to the peculiar alignment of the large-scale structure in the Local Universe, such as the Local Void, the Local Supercluster (LSC), and farther away structures like the Great Attractor (GA) and Perseus-Pisces cluster. This also includes studies of the Local Universe's velocity field, since it directly traces the gravitational potential and therefore the total matter distribution.

The ansatz of the CLUES project\footnote{www.clues-project.org}, which provides the framework for the study presented here, is to conduct \emph{constrained simulations} that are able to reproduce the LSS of the observed Local Universe \citep{Klypin2003,Gottloeber2010arXiv}. Such simulations serve as an ideal laboratory for studying structure formation in our cosmological neighbourhood \citep{Libeskind2010,Forero2011,Knebe2011a,Knebe2011b}. They are constructed by constraining the initial conditions using observational data \citep{Hoffman1991,Bistolas1998,Zaroubi1995,Zaroubi1999}, in particular we will use the measured radial peculiar velocities of galaxies in the Local Universe \citep{Tully2008,Tully2009,Courtois2012} as the input data. To construct initial conditions for these simulations, the constrained realizations (CR) algorithm \citep{Hoffman1991} is used, which combines a Bayesian reconstruction from the data with a random component created with a conventional initial conditions generator.

From previous CLUES simulations \citep{Klypin2003, Gottloeber2010arXiv} we know that the technique of using peculiar velocities as constraints has its limits. These simulations used radial velocities from the now outdated MARK III \citep{Willick1997}, SBF \citep{Tonry2001}, and \citet{Karachentsev2004} catalogues. To obtain reasonable reproductions of the Local Universe's structure, these velocity constraints needed to be complemented by additional density constraints, which were drawn from X-ray selected cluster data \citep{Reiprich2002}. But even in this case only the few most massive clusters of the Local Universe, namely the LSC, the GA, and for larger boxsizes the Coma and Perseus-Pisces superclusters, were robust features appearing in the constrained realizations, with smaller scales essentially unconstrained and dominated by the random component. It is important to understand whether these limitations stem from insufficient accuracy of the observational data used as constraints, the method employed to construct the constrained initial conditions, or some fundamental physical limitation. Through a collaboration of CLUES with the observational Cosmicflows program \citep{Courtois2011a,Courtois2011b,Courtois2012arXiv,Tully2012arXiv}, it will become possible to use significantly higher-quality data with many more peculiar velocity datapoints out to higher distances, with better sky coverage and much smaller observational errors, for generating constrained simulations.  Here, we want to investigate how much increase in accuracy we can expect from constrained simulations set up with this new data compared to what we have now, and how the method itself of setting up constrained initial conditions can be improved in order to optimally utilize the additional information contained in the data. With these studies, we hope to pave the way for a new generation of accurate constrained simulations that will be produced by our collaboration during the next years, to provide a powerful framework in which the Local Universe can be studied in detail.

This work is the third in a series of papers on this subject. In the first paper (Doumler et al. 2012a, from here on Paper I) 
we presented the Reverse Zeldovich Approximation (RZA) method, which significantly increases the quality of reconstructed initial conditions obtained from peculiar velocity data. This is accomplished by a Lagrangian reconstruction of the primordial density distribution underlying the observed velocity field, essentially shifting the data back in time and thus provide better constraints for initial conditions than the original dataset observed at $z=0$. 
In the second paper (Doumler et al. 2012b, from here on Paper II) 
we then investigated the impact of observational errors on the RZA method. Here, we study how well constrained simulations can reproduce the underlying universe, if their initial conditions were constructed by employing the RZA method. For this we set up a detailed test by using realistic mock peculiar velocity data drawn from a test simulation snapshot at $z=0$. We use the data to generate constrained initial conditions at some early initial redshift $z_\tm{init}$ and run them forward again with an $N$-body code. This evolved re-simulation is then compared to the original simulation at $z=0$. We do this for both the previous CLUES method and our new RZA method to compare by how much the simulation accuracy improves by performing Lagrangian reconstruction on the data. We also compare two mocks of different quality, to estimate how much is gained by using more accurate data.

The outline of this paper is as follows. In Section \ref{sec:method}, we briefly review our method of generating constrained initial conditions from the data, describe the setup of this test, and present the set of re-simulations we conducted. In Section \ref{sec:results}, we study the accuracy of these re-simulations compared to the original reference simulation and present our findings. We summarize and discuss our results in Section \ref{sec:summary}.

\section{Method and test setup}
\label{sec:method}

\subsection{Initial conditions with RZA}
\label{subsec:method2}

Our method of RZA reconstruction and subsequent generation of constrained initial conditions is described in detail in Paper I; we give only a brief summary here.

We start with a set of datapoints, i.e.\ radial peculiar velocities $v_r$ at discrete positions $\bl r$ at $z=0$. We first apply a grouping procedure to the data in order to ``linearize'' it, i.e.\ to remove virial motions and other small-scale interactions inside galaxy groups and clusters and to produce a data set that traces the coherent large-scale velocity field. We then reconstruct the three-dimensional peculiar velocities $\bl v (\bl r)$ at positions $\bl r$ by using the Wiener Filter (WF). The WF produces an estimate $\bl v^\tm{WF} (\bl r)$ based on the correlation function given by the assumed prior model, which is defined through the cosmological parameters and power spectrum $P(k)$. It also filters out noise due to observational errors from the data. 
In order to construct cosmological initial conditions at some early redshift $z_\tm{init}$, we need to obtain a suitable set of constraints. We do this with the RZA method: we estimate the initial position $\bl x_\tm{init}$ at $z_\tm{init}$ of our peculiar velocity field tracers, which are located at $\bl r$ at $z = 0$, with
\begin{align}
\label{eq:xinit_rza}
\bl x_{\tm{init}}^{\tm{RZA}}  = \bl r - \bl \psi^{\tm{RZA}}\;\;\; = \bl r - \frac{\bl v^\tm{WF}}{H_0 f}\;\;\;.
\end{align} 
The reconstructed initial positions $\bl x_\tm{init}$ are not to be understood as actual positions of the observed galaxies at early times --  at typical initial redshifts $z_\tm{init}$, no galaxies have formed yet. They are rather interpreted as a way to trace the estimated initial peculiar velocity field at $z_\tm{init}$. In order to construct the full set of constrained initial conditions, we then shift the original datapoints $v_r$ ``back in time'' to $\bl x_\tm{init}$ and use them as constraints to construct a constrained realization with the \citet{Hoffman1991} method. We only change the position of the constraints from $\bl r$ to $\bl x_\tm{init}$, but preserve the amplitude of the velocity and the direction of the component that is constrained (note that it is not the radial direction anymore w.r.t. the observer because the position has changed). The constrained realization is then obtained by evaluating
\begin{align}
\label{eq:deltacr}
\delta^\tm{CR} (\bl r) = \delta^\tm{RR} (\bl r) + \langle \delta(\bl r) c_i \rangle \,\langle c_i c_j \rangle^{-1}\, (c_j - \tilde{c}_j) \;\;\;,
\end{align}
where $\delta^\tm{RR}$ is an independently generated random realization, $c_i$ are the RZA-shifted datapoints, $\tilde{c}_i$ are the corresponding values of the same quantities in the random realization, and the angled brackets denote the values of the correlation functions of the different quantities, defined by $P(k)$. In order to construct an actual set of initial conditions for an $N$-body simulation, we scale $\delta^\tm{RR}$ to $z_\tm{init}$ and solve for the displacement field, $\bl \psi (\bl r) = - \bl \nabla^{-1} \delta(\bl r)$. We then place particles on a grid with displacements $\bl \psi$ according to the established Zeldovich-approximation method \citep{Efstathiou1985}.

\subsection{Mock data}

\begin{figure}
\centering
\includegraphics[scale=1.0]{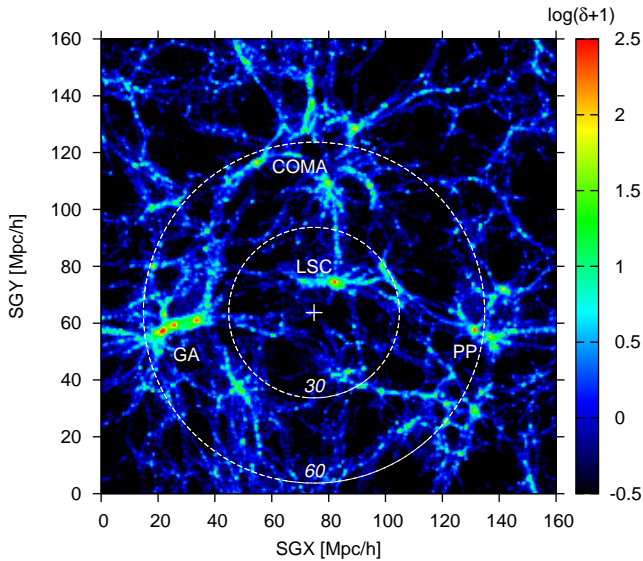}
\caption{Slice through the BOX160 constrained simulation from where
the mock catalogues have been extracted. The cross is placed at the position
$\bl r_\tm{MW}$ of the mock observer. The dashed circles illustrate the mock data
volume for mocks with Rmax = 30 Mpc/$h$ and 60 Mpc/$h$, respectively.}
\label{fig:simumap_paper2}
\end{figure}

As we already did in Papers I and II, we use the BOX160 simulation conducted by the CLUES project as the source of our mock galaxy peculiar velocity data. Again, we refer the reader to these papers for details. The BOX160 is a constrained simulation of the Local Universe with a boxsize of 160 Mpc/$h$, set up with the WMAP3 cosmological parameters. The simulation contains a large-scale structure resembling the observed Local Universe (see Figure \ref{fig:simumap_paper2} and \cite{Cuesta2011}). The simulation contains a configuration of objects corresponding to the LSC, GA, Coma, and Perseus-Pisces clusters (labelled in Figure \ref{fig:simumap_paper2}), and a Local Group (LG) candidate. We select the position of this LG object (marked as a white cross in Figure \ref{fig:simumap_paper2}) as the mock observer and generate from there realistic mock observational catalogues of galaxy peculiar velocities. A detailed description of how the mocks are built can be found in Paper I. Here we want to mention again that the mocks realistically reproduce features of real observational catalogues, such as a limited distance, knowledge of only the radial component $v_r$ of $\bl v$, sparse sampling, and observational errors due to inaccurate galaxy distance measurements. In this paper, we concentrate on the particular mocks C30\_10 and E60\_10, which are designed to mimic the current Cosmicflows-1 catalogue and the upcoming Cosmicflows-2 catalogue, respectively. After grouping, the catalogue C30\_10 contains 588 radial peculiar velocity datapoints within a relatively small radius of 30 Mpc/$h$ from the mock observer. This is a similar quality like the datasets that were previously used to construct constrained simulations. The E60\_10, on the other hand, contains 7632 datapoints within 60 Mpc/$h$, and models the quality of the upcoming new Cosmicflows-2 dataset, which we plan to use for constrained simulations of the Local Universe in future work. The data zone radii of 30 and 60 Mpc/$h$, respectively, are marked in Figure \ref{fig:simumap_paper2} with dashed white circles.

Since we want to study specifically how well realizations of cosmological initial conditions can be constrained with peculiar velocity data, and how the RZA method performs in this context, in this work we do not use any other types of constraints such as cluster density constraints for our re-simulations.


\subsection{The set of constrained realizations}

For each of the two mocks, C30\_10 and E60\_10, we construct several constrained realizations of cosmological initial conditions. We use the method outlined in Section \ref{subsec:method2}, which combines the CR method with RZA reconstruction. In the following, we refer to this procedure as ``Method II''. Furthermore, we also generate initial conditions with the method previously used for CLUES simulations \citep{Klypin2003,Gottloeber2010arXiv}, which we call here ``Method I''. It consists of using the peculiar velocity datapoints at $z=0$ directly as constraints for Eq. (\ref{eq:deltacr}), omitting the RZA shift. The main 
drawback of Method I is that it treats the peculiar velocities as though linear theory would be valid at all scales at $z=0$, neglecting all higher-order effects such as the cosmological displacement field $\bl \psi$. This leads not only to a poorer reconstruction quality, but also to a systematic position error of the clusters recovered in such simulations (compared to their observed counterparts). Typically, the object's positions will be off by the amplitude of $\bl \psi$, which is about 10 Mpc/$h$ on average at $z=0$. These shifts were observed in all previous CLUES simulations. We showed in Paper I that RZA can compensate for the shifts; here, we want to demonstrate how this improved method affects the outcome of evolved constrained simulations at $z=0$.

Having two different mock catalogues and two different methods to generate ICs, we also want to study the impact of the random component $\delta^\tm{RR}$ in Eq. (\ref{eq:deltacr}). The peculiar velocity constraints $c_i$ affect only large scales from $\approx 5$ Mpc/$h$ upwards, and only in regions of the box well covered by the data; all other structures emerging in the constrained simulation will have their origin in the particular realization of the random component. We therefore expect that the random seed has a large impact on the outcome of the simulation. Varying the seed while keeping all other parameters such as the constraints constant allows us to estimate how robustly structures that are constrained by the data are actually recovered in the constrained simulations.

For each mock-method combination, we created six different realizations with different seeds for the random component. We have therefore a set of 24 different realizations of initial conditions. 
We use this set to test our method on scales smaller than the box. In all cosmological simulations - constrained as well as unconstrained - due to periodic boundary conditions the dynamics on the scale of the box is incorrect. Therefore, one must expect that also in future constrained simulations based on observational data the bulk flow on scales of the order of the simulation box will be incorrect.
We construct the initial conditions on a regular cubic grid with a resolution of $N=256^3$ and a boxsize of $L=160$ Mpc/$h$, matching the boxsize of the ``source simulation'' BOX160\footnote{The original BOX160 simulation was performed with the ART code \citep{Kravtsov1997} and a resolution of $N=1024^3$. However, since it is impractical to run a large number of re-simulations with such a resolution, we limited ourselves to $N=256^3$ and also re-ran the BOX160 with Gadget-2 and this lower resolution to serve as the reference simulation. This way we can stay fully consistent on the parameters of all simulations.}.
All realizations were constructed with our newly developed numerical code \textsc{ICeCoRe} (see Paper I for details). For the 24 different realizations, we assumed the following naming convention. We abbreviate the six different seeds as A through F. We then add the method and seed numbers to the end of the mock name, so that for example C30\_10\_II\_A refers to the first out of six realizations that were constructed with constraints from the C30\_10 mock using Method II. We then ran each of the generated constrained realizations of initial conditions forward until $z=0$ with the simulation code Gadget-2 \citep{Springel2005a}, using collisionless particles only (no SPH particles) with a resolution of $N=256^3$ particles. We also used the same cosmological parameters and initial power spectrum $P(k)$ that was used for the BOX160 simulation, to be fully consistent on the assumed cosmological model. 
In fact, assuming a different cosmology would change the result.  For example, increasing substantially the normalization of the power spectrum without changing the constraints leads to a faster evolution. Instead of a local group like object one would find at the same place  a massive large halo of about the total mass of the group.

\section{Results}
\label{sec:results}

\subsection{Scatter and mass function}

\begin{figure}
\centering
\includegraphics[trim=2.5cm 0cm 2cm 0cm, clip=true, scale=0.73]{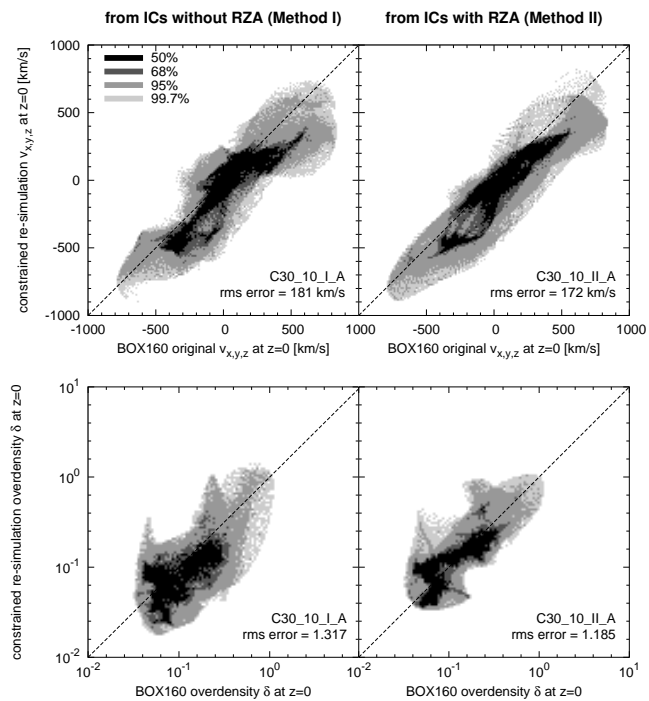}
\caption{Cell-to-cell comparison between the evolved constrained resimulations at $z=0$ for and the original BOX160 simulations. Top row: velocity field inside the mock volume (all three components were concatenated); Bottom row: density field. All fields were smoothed with a 5 Mpc/$h$ Gaussian.}
\label{fig:resimscatter_C30}
\end{figure}

\begin{figure}
\centering
\includegraphics[trim=2.5cm 0cm 2cm 0cm, clip=true, scale=0.73]{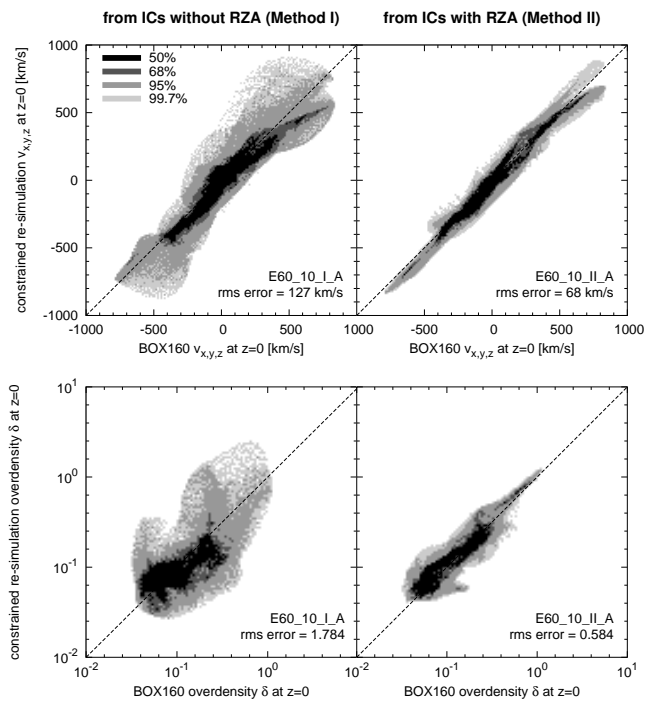}
\caption{As Figure \ref{fig:resimscatter_C30}, but for constrained realizations from the E60\_10 mock with twice the data zone radius.}
\label{fig:resimscatter_E60}
\end{figure}

\begin{figure*}
\centering
\includegraphics[scale=1.1]{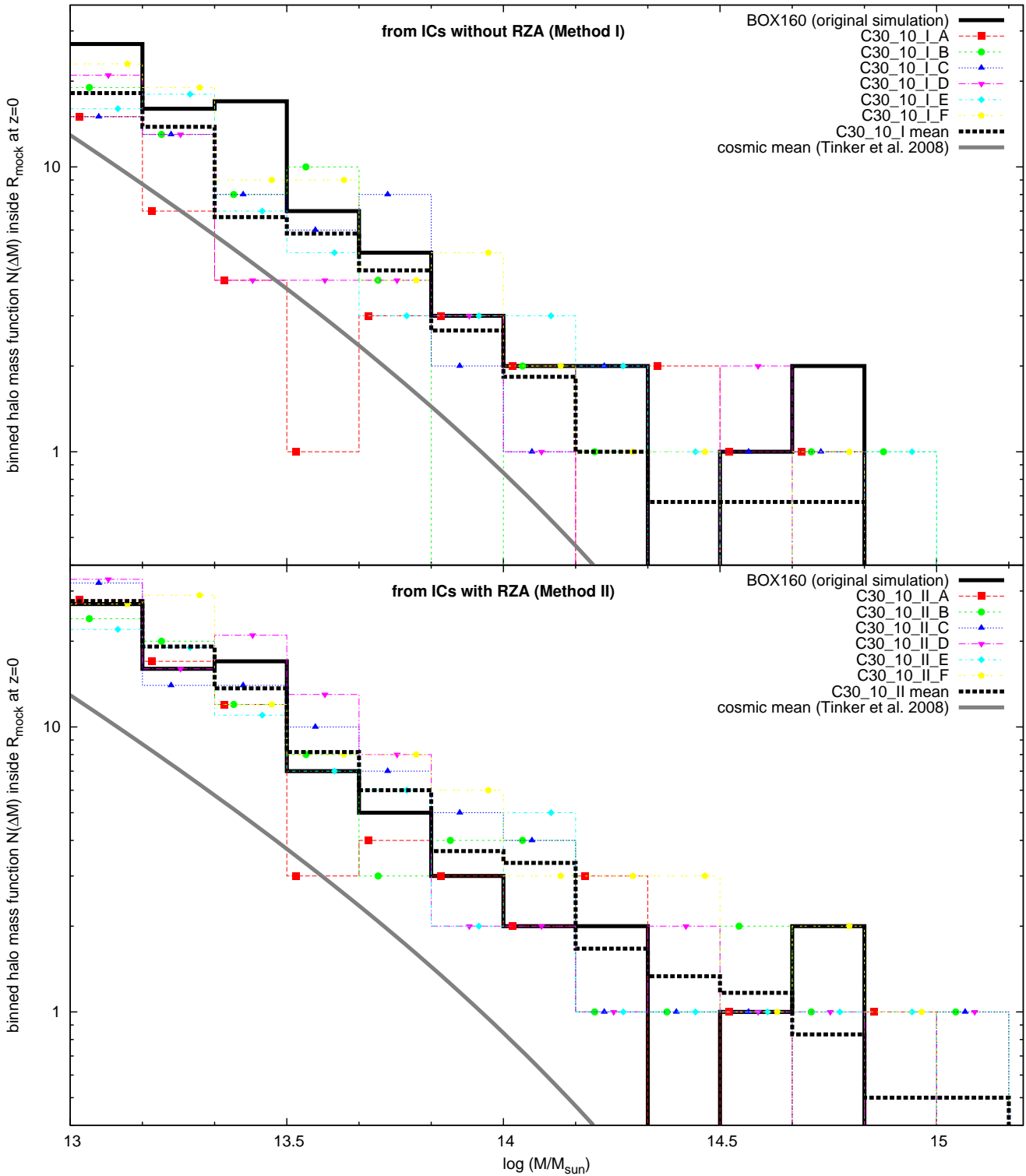}
\caption{Dark matter halo mass function inside the 30 Mpc/$h$ mock volume for the six different realizations from C30\_10 without RZA (top) and with RZA (bottom). The individual realizations are shown with coloured lines/points; their average mass function is the thick black dashed line. The original mass function of the BOX160 in this volume is the thick solid black line. The mass function for an average subvolume of radius 30 Mpc/$h$ with the BOX160 parameters is shown in grey (theoretical model of \citealt{Tinker2008}).}
\label{fig:massfunc}
\end{figure*}

Figure \ref{fig:resimscatter_C30} shows a cell-to-cell comparison between the evolved constrained resimulations obtained from the C30\_10 mock and the original field of BOX160 in the constrained volume (out to 30 Mpc/$h$ distance from the mock observer), both for Method I (without RZA) and Method II (with RZA). Figure \ref{fig:resimscatter_E60} displays the same for the resimulations from the E60\_10 mock. It can be seen that all mocks and methods produce density and velocity fields that are well-correlated with the original BOX160 universe at $z=0$. However, especially for the more accurate E60\_10 mock, the added RZA reconstruction that was used to obtain the initial conditions in Method II significantly improves the correlation over the one obtained with Method I (without RZA). For the E60\_10 mock, the peculiar velocity field is reproduced accurately down to an accuracy of about 1/4 of its total standard deviation with Method II that uses RZA. This is an impressive improvement of the reconstruction quality. From the correlation of the density fields, it can be seen that Method I, which does not use RZA, significantly underestimates the total density of the constrained region, which is overdense compared to the cosmic mean. The RZA method, on the other hand, accurately reproduces this total overdensity, for both the C30\_10 and E60\_10 resimulations. The total overdensity as well affects the abundance of dark matter haloes. Figure \ref{fig:massfunc} shows the binned dark matter halo mass functions of all realizations obtained from the C30\_10 mock catalogue, both with and without RZA. The mass functions of the individual realizations are shown with coloured lines and points, their average is represented by the thick dashed black line, and the actual mass function of the BOX160 inside the 30 Mpc/$h$ region is shown with the solid dashed black line. This is an overdense region in the BOX160 simulation: the mass function of this region lies significantly above the cosmic mean (grey line). The volume contains about 2.5 times more haloes with masses $> 10^{13} M_\odot/h$ than a region of mean cosmic density would with the given cosmological model and boxsize. The constrained resimulations without RZA only partially follow that behaviour. Their average mass function is also above the cosmic mean, but still systematically underestimates the actual halo abundance in the corresponding region of the original BOX160 simulation. On the other hand, the resimulations where RZA was used for constructing the initial conditions follow the mass function of BOX160 much more closely.

\subsection{The BOX160 universe}

\begin{figure*}
\centering
\includegraphics[scale=0.95]{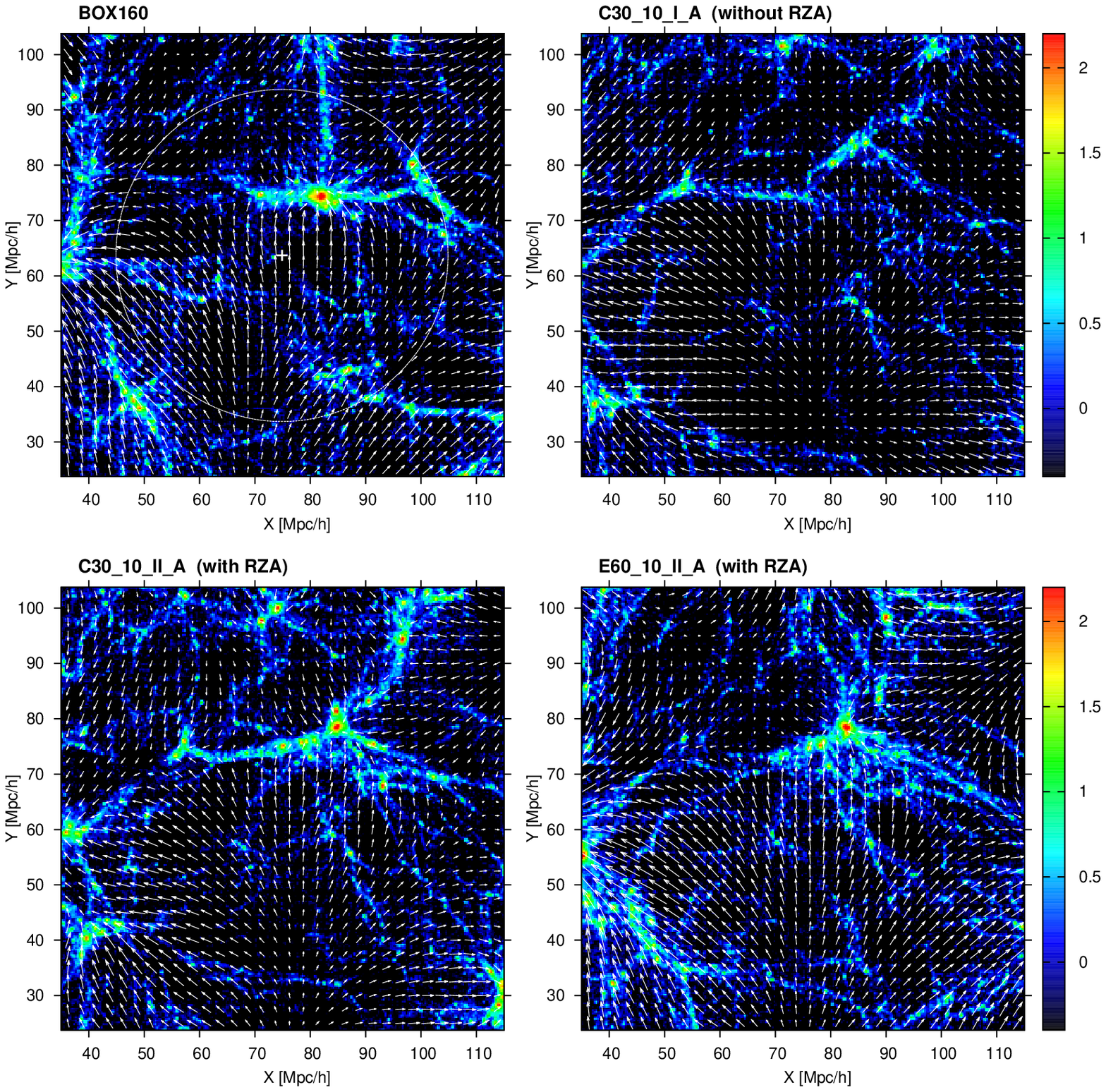}
\caption{Density and velocity fields in a 10 Mpc/$h$ thick slice at $87 < Z < 97$ Mpc/$h$ for the original BOX160 simulation (top left) and three constrained resimulations without RZA (top right), with RZA (bottom left), and with RZA from the E60\_10 mock using a larger data volume (bottom right). The white circle marks the constrained region for the C30\_10 mock. The density is shown in logarithmic scale. The arrows are proportional to the amplitude of the peculiar velocity at each position. The prominent object inside the BOX160 data zone is the simulated Virgo cluster with a virial mass of $3.25 \times 10^{14} M_\odot/h$. In all three constrained resimulations the same random seed was used.}
\label{fig:resimexamplemap}
\end{figure*}

The large-scale structure of the ``local universe'' for our test case, i.e.\ the region around the mock observer in the BOX160 simulation, is shaped by several dominant structures that to some degree resemble the observed Local Universe. The ``Milky Way'' analogue halo that was chosen as the position of the mock observer ($X=75; \, Y=64; \, Z=80$) lies in the vicinity of a ``Virgo'' cluster with a virial mass of $3.25 \times 10^{14} M_\odot$, which is embedded in a thick filament parallel to the $X$-axis. The local flow of the mock observer is directed towards this structure, resembling the observed Virgocentric infall \citep{Tully2008}. On a larger scale, the whole region is dominated by a flow towards the ``Great attractor'' (GA), a massive structure at about $X=30$, $Y=60$ that lies outside of the 30 Mpc/$h$ data zone of the C30\_10 mock. This configuration can be seen in the top left map of Figure \ref{fig:resimexamplemap}, which is a zoom-in from Figure \ref{fig:simumap_paper2}. The BOX160 Virgo cluster is however not the most massive structure within 30 Mpc/$h$, as there is an even more massive region at a distance of slightly less than 30 Mpc/$h$, lying in a direction in between Virgo and the GA, that contains two clusters with masses of $6.06 \times 10^{14} M_\odot/h$ and $5.20 \times 10^{14} M_\odot/h$, respectively (their $Z$ positions are just outside of the slice plotted in Figure \ref{fig:resimexamplemap}). We associate them with the Centaurus and Hydra clusters. They in turn cause a significant infall flow towards them on the surrounding structure. In the other direction (towards negative $Y$), there is also a large void that contributes to the shape of the large-scale velocity field by creating a push outwards of it, although this particular feature in the BOX160 may not be as dominant as the observed Local Void \citep{Tully2008}. Constrained resimulations of this test universe should be able to recover all these characteristic structures. BOX160 contains this specific configuration because as already mentioned it is itself a constrained simulation of the Local Universe. It is not entirely accurate, for example the virial mass of the BOX160 Virgo cluster is 2 -- 3 times less than the estimated virial mass of its observed counterpart \citep{Fouque2001}, but the described main characteristics of the large-scale structure are present. Ignoring the origin of BOX160 for this test and taking its large-scale matter distribution as ``given'', we try to reproduce it in the constrained resimulations, which should retain the main characteristics that BOX160 shares with the observed Local Universe in this ``second pass'' of the reconstruction-resimulation cycle.

\subsection{The re-simulated Virgo cluster}

The other panels of Figure \ref{fig:resimexamplemap} show three constrained realizations obtained from the C30\_10 mock catalogue without RZA (top right), with RZA (bottom left), and from the larger-volume E60\_10 mock catalogue with RZA, in the 10 Mpc/$h$ thick slice that should contain the BOX160 Virgo cluster. One can immediately notice that, while all three simulations faithfully reproduce the large-scale flow towards the GA, the resimulation that was constructed without RZA fails to create a Virgo cluster. Apparently, the mass of the BOX160 Virgo cluster of $3.25 \times 10^{14} M_\odot/h$ lies below the scale that could be reproduced with the non-RZA Method I of generating constrained initial conditions. Why then does the BOX160, which was created with the same method, have a Virgo cluster in the first place? 
There are three reasons: First, the observed Virgo cluster has a mass that is 2 -- 3 times higher, which places it in the recoverable mass range. Second, we observe a scatter in the recovered masses and BOX160 is a  realisation with high Virgo mass. Third, in the BOX160, the Virgo cluster was additionally constrained by density constraints placed on its observed position.  In our setup, we do not use such additional constraints.
The non-RZA resimulations show some overdensity in that region, and there is also the tendency of a flow in that direction, but there is not enough overdensity to form a cluster of comparable mass. On the other hand, both resimulations in Figure \ref{fig:resimexamplemap} that include RZA reconstruction have a cluster at that position, which is similarly embedded in a thick filament, with a strong flow towards it from the position of the mock observer.

\begin{table*}
\begin{center}
\begin{tabular}{cccccc}
\hline \hline \\[-3mm]
Simulation & Position $X, Y, Z$ & pec. vel. $v_x, v_y, v_z$ & Mass & Position error & Mass relative to \\
& [Mpc/$h$] & [km/s] & [$10^{14} M_\odot/h$] & [Mpc/$h$] & BOX160 Virgo \\[2mm]
\hline \\[-3mm]
BOX160 Virgo     & $82.0, 74.4, 91.8$ & $ +22, +142, -473$ & 3.25 & -- & -- \\ [2mm]
\hline \\[-3mm]

C30\_10\_II\_A & $84.8, 78.5, 92.6$ & $ -23,  -61, -391$ & 1.66 & 5.0 &  51 \% \\
C30\_10\_II\_B & $80.5, 76.8, 96.7$ & $ -22,   +6, -912$ & 1.98 & 5.7 &  61 \% \\
C30\_10\_II\_C & $80.3, 72.6, 94.4$ & $ +26, +135, -403$ & 3.24 & 3.6 &  100 \% \\
C30\_10\_II\_D & $83.6, 73.8, 92.1$ & $+101, +231, -326$ & 1.21 & 1.7 &  37 \% \\
C30\_10\_II\_E & $80.2, 76.5, 94.3$ & $ +72, +247, -755$ & 2.33 & 3.7 &  71 \% \\
C30\_10\_II\_F & $87.1, 78.0, 92.1$ & $-273,  +75, -527$ & 2.83 & 6.3 &  87 \% \\ [2mm]

C30\_10\_II mean & $82.8, 76.0, 93.7$ & $-19, +106, -552$ & 2.21 & 2.6  & 68 \% \\ 

standard deviation&  $2.6, 2.1,  1.6 $& $122, 112,  212$&  0.68 & 4.6 & 21\%\\ [2mm]
\hline \\[-3mm]

E60\_10\_II\_A & $82.8, 78.5, 90.5$ & $-113,  -64, -507$ & 1.87 & 4.4 & 58 \% \\
E60\_10\_II\_B & $78.6, 74.7, 93.9$ & $-200, +149, -682$ & 2.73 & 4.1 & 84 \% \\
E60\_10\_II\_C & $77.1, 72.8, 91.8$ & $ +29, +135, -374$ & 1.21 & 5.2 & 37 \% \\
E60\_10\_II\_D & $82.0, 76.0, 89.5$ & $-142, +201, -576$ & 3.03 & 2.8 & 93 \% \\
E60\_10\_II\_E & $78.5, 75.3, 92.2$ & $ +85, +289, -706$ & 1.31 & 3.7 & 40 \% \\
E60\_10\_II\_F & $84.0, 76.3, 88.8$ & $ +21, +259, -486$ & 2.80 & 4.1 & 86 \% \\ [2mm]

E60\_10\_II mean & $80.5, 75.6, 91.1$ & $  -53, +161, -555$ & 2.16 & 2.1 & 66 \% \\ 

standard deviation & $2.5,  1.7, 1.7$& $ 103, 115, 115 $& 0.73 & 4.1 & 22\% \\ [2mm]
\hline \hline
\end{tabular}
\end{center}
\caption{Virgo candidates found in the RZA resimulations of BOX160 at $z=0$. The first line corresponds to the original ``Virgo'' object in the BOX160. The following lines list the haloes found in the AHF catalogues of the respective resmulations that are within 10 Mpc/$h$ of the BOX160 Virgo position and have a virial mass of at least $10^{14} M_\odot/h$. 
}
\label{table:virgocandidates}
\end{table*}

In order to understand more clearly how robustly the Virgo cluster is recovered in the constrained resimulations we searched for it in all resimulations by searching for haloes that would be within 10 Mpc/$h$ of the original BOX160 Virgo's position and would have a mass of at least a $10^{14} M_\odot/h$. The result was that in all realizations using RZA, both from the C30\_10 mock and the E60\_10 mock, we could find a cluster corresponding to Virgo. These objects are listed in Table \ref{table:virgocandidates}. On the other hand, we could not find such an object in any of the realizations created without RZA, which clearly displays that the RZA improves recovering the original structures in constrained simulations. The resimulated Virgo is not exactly at its BOX160 position in the resimulations; the error lies between 1.7 and 6.3 Mpc/$h$ and varies with the random seed. The average position of all resimulated Virgos is only about 2 Mpc/$h$ away from its original position in BOX160, so this fluctuation is probably due to the influence of the random modes rather than a systematic shift. It is interesting to note that in the RZA resimulations, all the Virgo reproductions have a lower mass than the original BOX160 Virgo, just as the latter has a lower mass than the observed Virgo cluster. This may be connected to the findings of \citet{Courtois2012}, who analysed the Cosmicflows-1 distance catalogue with the Wiener filter and found that the Virgo cluster is not dominating the peculiar velocity field as much as expected. It may therefore be harder to recover accurately in a constrained simulation.

\begin{figure*}
\centering
\includegraphics[scale=1.015]{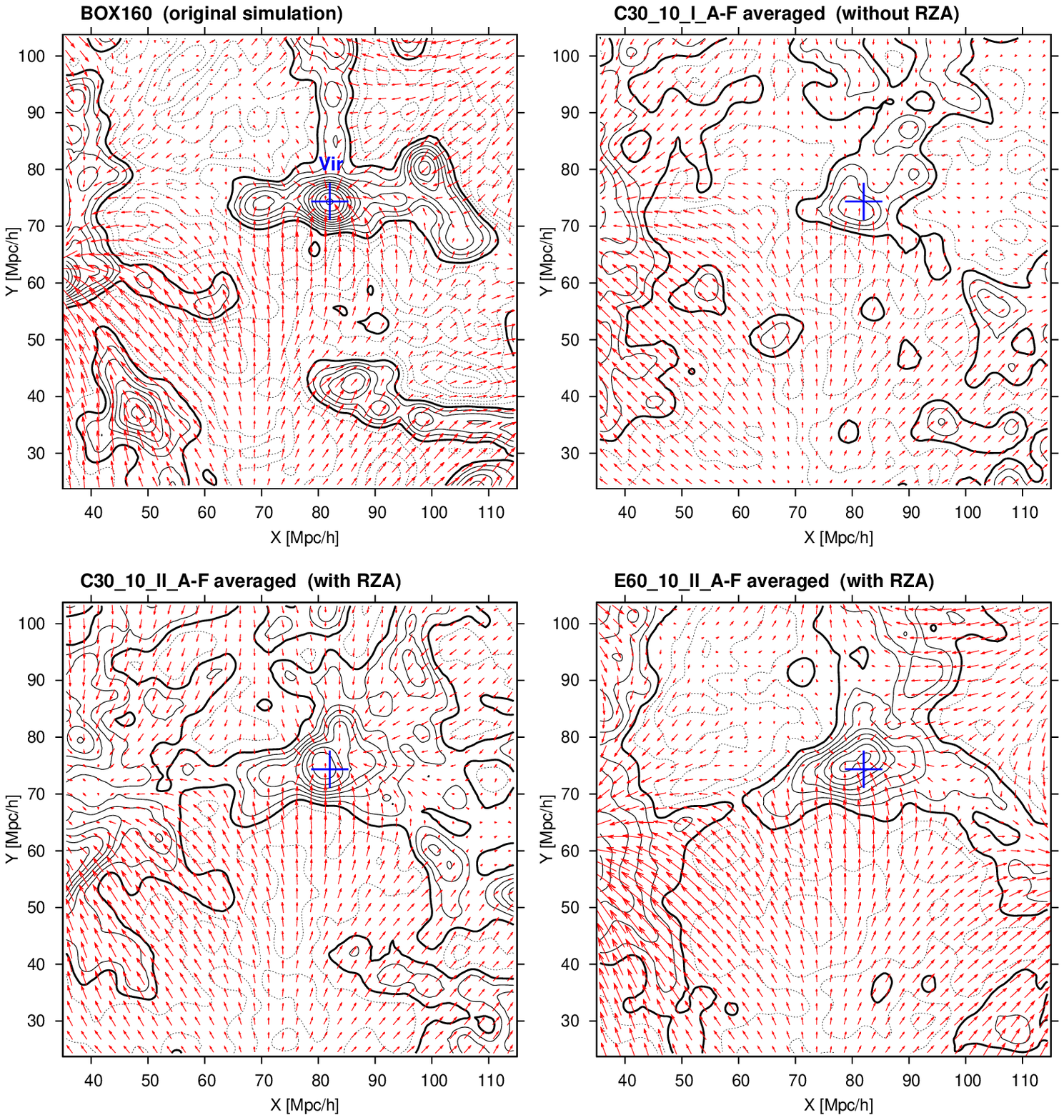}
\caption{Density and velocity fields in a 10 Mpc/$h$ thick slice at $87 < Z < 97$ Mpc/$h$ for the original BOX160 simulation (top left) and the average of all six constrained realizations without RZA (top right), with RZA (bottom left), and with RZA from the E60\_10 mock using a larger data volume (bottom right). The density fields were smoothed with a Gaussian of radius 2.5 Mpc/$h$. The original centre of the BOX160's Virgo cluster is marked with a blue cross in each map. The thick contour line marks the cosmic mean density; solid contour lines are drawn for overdensities and dotted contour lines for underdensities.}
\label{fig:averagedmap_vir}
\end{figure*}

Another possibility to estimate the robustness of the reconstruction is to compute an average of the density and velocity fields over the different evolved realizations. This way, structures coming from the random component will tend to average out and be suppressed, while structures appearing consistently in every realization will be enhanced. Figure \ref{fig:averagedmap_vir} shows the same slice as in Figure \ref{fig:resimexamplemap}, but averaged over all six realizations A -- F for the C30\_10\_I (without RZA), C30\_10\_II (with RZA), and E60\_10\_II (with RZA, larger mock). The position of the original BOX160 Virgo cluster is marked with a blue cross. The resimulations done without RZA (top right) show some overdensity in this region, and a tendency of a flow towards it, but the overdensity is not high enough to create a cluster of mass comparable to Virgo in any of the non-RZA realizations. The density peak is much more pronounced in the simulations with RZA (bottom panels), which all have a massive object near to that location. We also see that other structures with less mass, such as the overdense region below Virgo around $X=90 \; Y=40$, are not present in the averaged fields, which means that they lie in a mass range that is too low to be recovered by the reconstruction of initial conditions for any of the methods/mock catalogues.

\begin{figure*}
\centering
\includegraphics[scale=1.015]{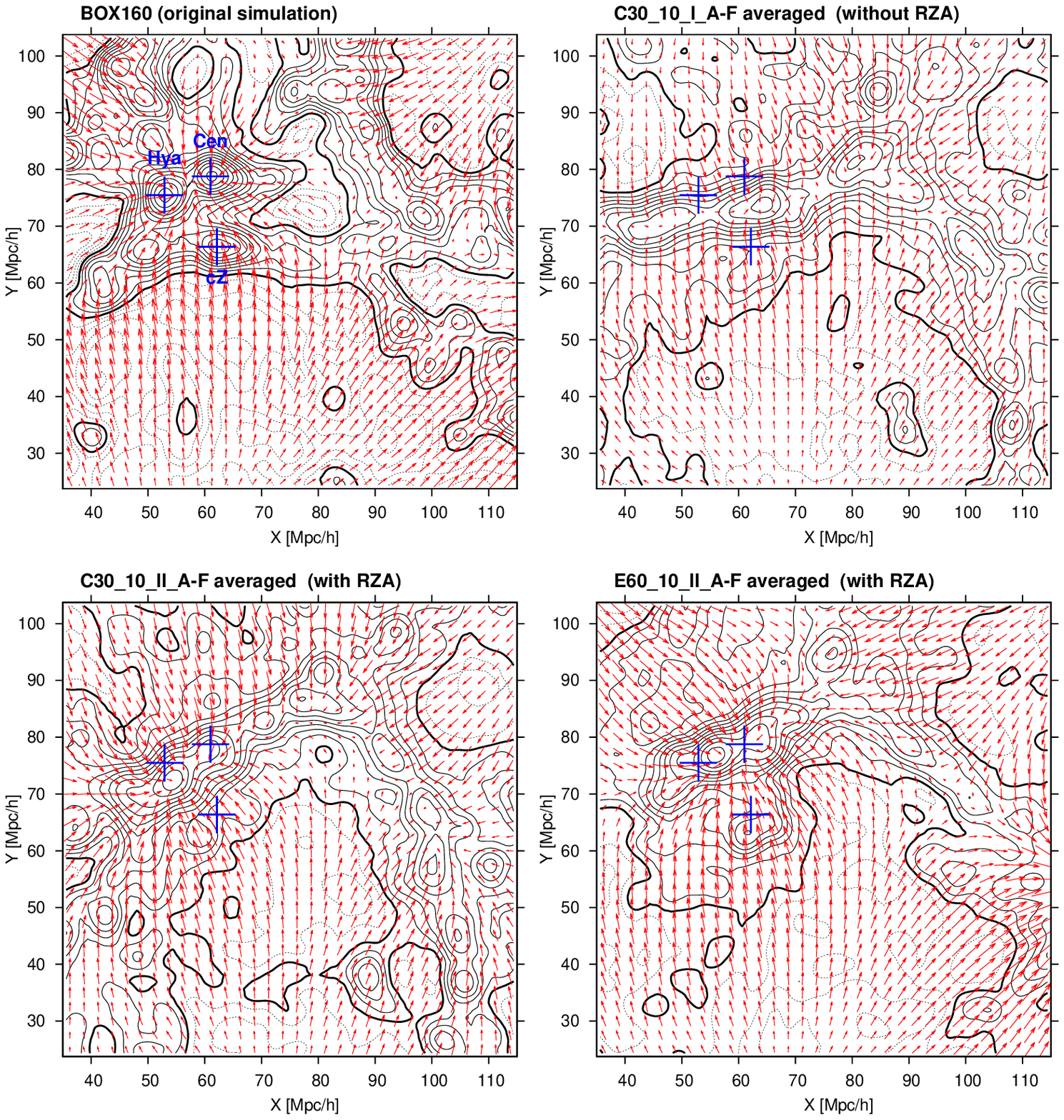}
\caption{Same as Figure \ref{fig:averagedmap_vir}, but for a different slice at $69 < Z < 79$ Mpc/$h$. In this slice, the BOX160 contains three massive clusters: Centaurus (Cen) with virial mass $6.06 \times 10^{14} M_\odot/h$, Hydra (Hya) with $5.20 \times 10^{14} M_\odot/h$, and ``Cluster Z'' (cZ) with $0.96 \times 10^{14} M_\odot/h$. The original centres of these three objects in BOX160 are marked with blue crosses in each map.}
\label{fig:averagedmap_hyacen}
\end{figure*}

\subsection{Positions and masses of other clusters}

To obtain a more precise estimate about the mass scale on which objects can be consistently recovered in the constrained resimulations using radial peculiar velocity constraints, we search for other massive clusters in the data zone. The largest overdense region within the data zone is dominated by the Centaurus and Hydra clusters, with masses of $6.06\times 10^{14} M_\odot/h$ and $5.20\times 10^{14} M_\odot/h$, respectively. These two clusters are separated by a distance of 10 Mpc/$h$. This overdense region is robustly recovered in all simulations including the ones from Method I not using RZA. It therefore lies above the minimum mass scale that can be recovered without RZA. Figure \ref{fig:averagedmap_hyacen} shows the corresponding slice of the density and velocity fields that contains the Hydra and Centaurus clusters, averaged over the different realizations\footnote{
Note that in the observational data the Hydra/Centaurus clusters and the Virgo cluster all lie within the supergalactic plane around $SGZ=0$ (see e.g.\ \citealt{Courtois2012}). On the other hand, while BOX160 uses the same coordinate system, there the Hydra/Centaurus clusters and the Virgo cluster are not located at the same $Z$ (=in the same $X/Y$ plane), but actually in planes about 10 -- 15 Mpc/$h$ apart from each other in the $Z$ dimension. This is another example of the systematic shifts that occur in constrained simulations from peculiar velocity data if one does not use Lagrangian reconstruction such as RZA.}. 
All resimulations show a massive overdensity in this region. However in the Method I resimulations there is a clear systematic shift in position. Further, in the averaged map one can see only one smeared out peak instead of two, which is additionally shifted towards positive $X$ and negative $Y$. In two of the six C30\_10\_I realizations (B and E), we can find only one massive cluster around the region where the Hydra-Centaurus pair should be. In these realizations, either the two overdense regions merged during their evolution due to shifts in their position and/or displacement, or already in the reconstructed initial conditions the accuracy was not sufficient to robustly resolve them into two separate peaks. On the other hand, in the RZA resimulations, both clusters are resolved robustly and show up as separate peaks that are very close to their intended position; in every realization using Method II, we can find appropriate objects for Hydra and Centaurus at $z=0$ within a few Mpc/$h$ of their original positions, just like for the Virgo cluster. Like in the case of Virgo, the masses of the resimulated Hydra/Centaurus clusters at $z=0$ show both a scatter and a systematic deviation from the original masses of the original BOX160 clusters of $6.06 \times 10^{14} M_\odot/h$ and $5.20 \times 10^{14} M_\odot/h$, respectively. The average mass of the resimulated Centaurus is at 90\% of the original mass in BOX160, with a scatter within a factor of two. Interestingly, the resimulated Hydra cluster is more massive than it should be, at 173\% of the original mass on average. In three out of the six C30\_10\_II realizations it is the more massive of the pair, although it should be the less massive, and in four out of six it even has a mass slightly above $10^{15} M_\odot/h$. As it is in the case of the Virgo cluster, this systematic error in the cluster masses does not improve if one goes from the C30\_10\_II realizations to the E60\_10\_II realizations.

If we add more constraints and increase the data volume, the mass scale that can be reproduced with the RZA method increases noticeably. For the C30\_10\_II simulations, we can always find a Virgo cluster, but already the next massive cluster (the fourth-massive halo within 30 Mpc/$h$) with $2.41 \times 10^{14} M_\odot/h$ cannot be unambiguously identified in one of the six realizations, the C30\_10\_II\_E. Namely, within a distance of 10 Mpc/$h$ and a mass within a factor of 3 there is no matching object in this realization. The next-massive clusters (ranked 5 to 7 by mass within 30 Mpc/$h$) cannot be reliably found anymore. On the other hand, if we go to the E60\_10\_II simulations, we find resimulated counterparts for all seven most massive clusters (the seventh having a mass of $0.96 \times 10^{14} M_\odot/h$) in all six realizations, all within a factor of 3 in mass and within 7 Mpc/$h$ in distance. The seventh most massive halo with $0.96 \times 10^{14} M_\odot/h$ appears as a clear peak in the averaged map of E60\_10\_II realizations (Figure \ref{fig:averagedmap_hyacen}, bottom right, labelled ``cZ'') close to its original position. For even lighter objects, it is not possible anymore to find unambiguous counterparts in the E60\_10\_II realizations. This is aggravated by the fact that below a certain mass there is a quickly increasing probability to find a seemingly matching, but actually randomly created object to appear around the right position.

The origin of the systematic errors on the reproduced objects' masses can be understood from the non-linearity of the structure formation process. The typical mass scatter within a factor of two that we observe here is consistent with the findings of \citet{Ludlow2011}, who studied the relation between virialised haloes and their protohalo peaks in the initial conditions. They found that a protohalo peak on some fixed scale determines the mass of the resulting halo only within a factor of two. This explains the cluster mass discrepancies in our constrained resimulations. The exact virial mass of an object at redshift $z=0$ is a product of various non-linear structure formation processes, such as at what rate it can accrete mass from the surrounding structure and how efficiently it is fed from outside by connected filaments. Such details of the structure around clusters cannot be recovered from reconstructed linear initial conditions. In the mass function of the 30 Mpc/$h$ data zone (Figure \ref{fig:massfunc}) we see that BOX160 has a very specific distribution: two objects (Hydra and Centaurus) in the highest-mass bin, one object (Virgo) in the next bin, and no objects in the bin after that. The average mass function of the constrained realizations does not follow this peculiar shape, but instead tends towards a distribution that is statistically more likely to occur.

\subsection{Filaments and voids}

Apart from massive dark matter haloes, another characteristic of the large-scale distribution of matter is the presence of filaments and voids. An example in the selected BOX160 data zone is the filament parallel to the $Y$-axis, which goes upward from Virgo towards the BOX160's Coma cluster (outside of the map); it is not recovered consistently in the RZA resimulations. We saw already in the RZA reconstruction of the displacement field (Paper I), that the RZA reconstruction struggles to recover such filamentary features accurately. This could be due to the non-linear structure formation in such regions: in general, structure formation first proceeds along one dimension, forming sheets, followed by a collapse in the next dimension, forming filaments, and finally the material in these filaments falls into haloes \citep{Shandarin1989}. The peculiar velocity field at $z=0$ only retains information about this last stage. In the particular case of the filament that connects the Virgo cluster to the Coma cluster in BOX160, the input data contains only the infall of objects along the filament toward Virgo, but not the initial velocity distribution that led to the formation of that filament, and therefore our reconstruction procedure (WF + RZA) cannot recover it. In the case of the C30\_10\_II\_A resimulation, this filament is instead replaced by another filament that is imposed by the random modes and aligned differently. This behaviour is repeatedly seen in the resimulations at other locations as well. At the other end of the filament there is a flow towards the Coma cluster seen at the upper edge of the slice in Figure \ref{fig:resimexamplemap} (around $X=80; \; Y>100$). This flow is missed by all resimulations that use the C30\_10 mock, since the BOX160 Coma cluster is too far away from the data zone. On the other hand, in the E60\_10\_II resimulations, this flow is recovered accurately because of the enlarged data zone, but this is not sufficient to also faithfully reproduce the alignment of the filament. 

It is not entirely clear at this point why the alignments of filaments are not always correctly reproduced in constrained simulations. Besides the inherent non-linearity of filament formation and therefore the inability to extrapolate this process back in time with RZA another possible reason could be the fact that the input data consist of only radial peculiar velocities and the accuracy of filament reconstruction could depend on how the filament is aligned compared with the line of sight.

We therefore cannot generally trust that the alignment of filamentary structure in the Local Universe can be sufficiently reproduced in constrained simulations, except if their formation is strongly constrained by massive objects inside the data zone, as is the case for the thick filament hosting the Virgo cluster. The presence of the latter is generally reproduced in all resimulations with RZA, although again its exact alignment is unconstrained.

We did not compare the alignment of voids in the resimulations in detail, but we see that the presence and alignment of the most prominent voids in the data zone is generally recovered well. The largest voids develop by expansion of underdensities in the initial conditions that are above the minimum scale of initial conditions reconstruction and are therefore sufficiently constrained. The reconstruction of the displacement field using RZA helps track the expansion of voids back in time and generate appropriate initial conditions for them. Since it is believed that in the Local Universe, the outward push caused by the Local Void has an important influence on shaping the Local Flow \citep{Tully2008}, it would be interesting to study in detail to what extent this behaviour can be seen in the constrained simulations. This will be the subject of future work.

\subsection{The re-simulated peculiar velocity field}

To conclude the analysis of our re-simulations, we want to mention the reproduction quality of the BOX160 peculiar velocity field at $z=0$. We see that within the datazone, the peculiar velocity field is recovered exceptionally well in the RZA resimulations, especially for the better E60\_10 mock. We already saw that in this case the deviation of the resimulations from the original BOX160 peculiar velocity field is only on the level of 1/4 of the total standard deviation of the field. Comparing in Figure \ref{fig:resimexamplemap} the peculiar velocity field of the original BOX160 (upper left) and the E60\_10\_II resimulation (bottom right), the structure of these two velocity fields is practically identical. The flows towards the major overdense regions and outwards of the large voids are all present with a correct alignment and amplitude. We can now argue that the limited ability to recover correct masses for particular clusters or the correct alignment of filaments plays a lesser role and does not substantially limit the overall meaningfulness of the constrained simulation, since the large-scale velocity field is recovered accurately by the constrained resimulations. If we imagine that a Local Group-like object would be placed at the centre of the E60\_10\_II resimulation in Figure \ref{fig:resimexamplemap}, then it would experience practically the same large-scale flow and would be embedded in a very similar large-scale environment compared to the original BOX160, so that we can study its dynamics. This is precisely the main motivation of running constrained simulations, where of course the reference universe is the actually observed Local Universe.


\section{Summary and conclusion}
\label{sec:summary}

In this paper, we investigated cosmological simulations that are designed to reproduce the large-scale structure of the Local Universe.  Such simulations can be produced by constraining their initial conditions with the constrained realizations (CR) algorithm of \citet{Hoffman1991}, using radial peculiar velocity data as input.  On top of the previously established CR method we have added the RZA reconstruction, which is a novel Lagrangian reconstruction scheme designed for peculiar velocity data. This new method allows to produce significantly more accurate constraints for the initial conditions. To quantify the accuracy of the method, we perform tests using mock data extracted from a previous simulation serving as the ``reference universe''. We use different mock catalogues with realistic observational errors that mimic the real data used for constrained simulations. After a reconstruction of the initial conditions from the mock data, we evolve them forward again until $z=0$ in a series of re-simulations and compare their outcome to the original reference simulation at $z=0$. We do so for both the previous method (without RZA) and our new method (with RZA). 

We find that with the previous method of generating constrained initial conditions, if we use a sparse mock limited to 30 Mpc/$h$ and do not employ a Lagrangian reconstruction scheme (as in previous constrained simulations within the CLUES project), the threshold for robustly recovering structures is around $\approx 5 \times 10^{14} M_\odot/h$, and the accuracy on their positions is typically at the scale of 10 Mpc/$h$. Adding the RZA reconstruction, we can lower this threshold to under $3 \times 10^{14} M_\odot/h$, enough to robustly recover an object in our test simulation that is similar to the Virgo cluster. If we use the more complete E60\_10 dataset, which covers a larger volume and approximates the accuracy of upcoming new datasets, the quality of reconstruction without using RZA does not increase much; but if we use RZA, we achieve a significantly better resolution that allows us to robustly recover structure down to mass scales of about $1 \times 10^{14} M_\odot/h$. Additionally, the RZA method reduces the errors on the positions where the resimulated structures appear to a scatter within about 5 Mpc/$h$ around the true position. The average position of the different realization is even only 2 Mpc/$h$ from the true position, meaning that there is practically no \emph{systematic} shift of structures; this is a significant improvement over previous constrained simulations from peculiar velocities, which featured shifts of the order of 10 Mpc/$h$ and more due to a failure to account for the cosmological displacement field.

We find that even with RZA and good input data, the method generally cannot reliably recover structures on mass scales below $\approx 1 \times 10^{14} M_\odot/h$ and also struggles to reproduce other large-scale features like the alignment of filaments. On the other hand, the peculiar velocity field at $z=0$ is reproduced exceptionally well by the re-simulations. This may be due to the fact that these velocities were used to constrain the realizations in the first place, and that they are less susceptible to non-linear effects than the cosmic density distribution. The ability to reproduce the peculiar velocity field makes constrained simulations constructed with the RZA method an ideal laboratory to study velocity flows in the Local Universe when applied to actual observational data. We expect that with real data, the same degree of accuracy can be obtained with our technique as we showed in the test presented here. Such an application of the method to  the newest observational peculiar velocity data is a main focus of our future work. We expect that these simulations will present a significant  methodical improvement over the currently available CLUES simulations and will provide a very useful framework for studying the dynamics of the Local Universe.


\section*{Acknowledgments}

TD would like to thank Steffen Knollmann for his help with computing the \citet{Tinker2008} mass function and generating Figure \ref{fig:massfunc}. YH and SG acknowledge support by DFG under GO 563/21-1. YH has been partially supported by the Israel Science Foundaction (13/08). TD and SG acknowledge support by DAAD for the collaboration with H.M. Courtois and R.B. Tully. We would like to thank the referee of this series of papers for her/his very careful and fast reading of the manuscripts and the many constructive comments which improved the three papers substantially.


\bibliography{Doumler2012_paper3} \bsp

\label{lastpage}

\newpage

\end{document}